\documentclass{article}
\usepackage[T1]{fontenc} %
\usepackage[utf8]{inputenc} %
\usepackage{ismir,amsmath,cite,url}
\usepackage{mathtools}
\usepackage{amssymb}
\usepackage{latexsym}
\usepackage{bm}
\usepackage{xfrac}
\usepackage[neveradjust]{paralist}
\usepackage{graphicx}
\usepackage{subcaption}
\usepackage{tikz}
\usepackage{color}
\usepackage{booktabs}
\usepackage[normalem]{ulem}
\usepackage{scrextend}
\usepackage{anyfontsize}
\usepackage{enumitem}
\usepackage{hyphenat}

\usepackage[bookmarks=false,hidelinks]{hyperref}

\usepackage[nameinlink,capitalize]{cleveref} %

\usetikzlibrary{calc,positioning,arrows,decorations.text}

\def\footurl#1{\footnote{\url{#1}}}
\def\clap#1{\hbox to 0pt{\hss #1\hss}}

\DeclarePairedDelimiter\abs{\lvert}{\rvert}

\newcommand{\timesig}[2]{\,\ensuremath{%
  \vcenter{\offinterlineskip
    \halign{\hfil##\hfil\cr
            $\scriptstyle\mathbf{#1}$\cr
            \noalign{\vskip0.16ex}
            $\scriptstyle\mathbf{#2}$\cr}
  }\,}%
}

\newcommand{\track}[1]{\textsc{#1}}
\newcommand{\trackpair}[2]{\track{#1}$\to$\track{#2}}

\title{Supervised Symbolic Music Style Translation Using Synthetic Data}

\multauthor
{Ond\v{r}ej C\'{i}fka \hspace{1cm} Umut \c{S}im\c{s}ekli \hspace{1cm} Ga\"{e}l Richard} {LTCI, Télécom Paris, Institut Polytechnique de Paris \\
{\{\texttt{ondrej.cifka}, \texttt{umut.simsekli}, \texttt{gael.richard}\}\texttt{@telecom-paris.fr}}
}
\def\authorname{Ond\v{r}ej C\'{i}fka, Umut \c{S}im\c{s}ekli, Ga\"{e}l Richard}

\begin{document}

\maketitle
\begin{abstract} 
Research on style transfer and domain translation has clearly demonstrated the ability of deep learning-based algorithms to manipulate images in terms of artistic style.
More recently, several attempts have been made to extend such approaches %
to music (both symbolic and audio) in order to enable transforming musical style in a similar manner. 
In this study, we focus on \emph{symbolic music} with the goal of altering the `style' of a piece while keeping its original `content'. 
As opposed to the current methods, which are inherently restricted to be unsupervised due to the lack of `aligned' data (i.e.\ the same musical piece played in multiple styles), we develop the first fully  \emph{supervised} algorithm for this task. At the core of our approach lies a \emph{synthetic data} generation scheme which allows us to produce virtually unlimited amounts of aligned data, and hence avoid the above issue. 
In view of this data generation scheme, we propose an encoder-decoder model for translating symbolic music \emph{accompaniments} between a number of different styles. %
Our experiments show that our models, although trained entirely on synthetic data, are capable of producing musically meaningful accompaniments even for real (non-synthetic) MIDI recordings.
\end{abstract}
\section{Introduction}
\label{sec:introduction}

Artistic style transfer has become a well-established topic in the computer vision literature and is becoming of increasing interest in other areas of computer science, especially music and natural language processing.
More generally, we are dealing with a family of \emph{style transformation} tasks, where the goal is to alter the \emph{style} of a piece of data (e.g., an image, a musical piece, a document) while preserving~-- to some extent~-- its \emph{content}.
In the music domain, a solution to these problems would have exciting industrial applications, not only as a way to generate new music automatically (as an alternative to fully automatic music composition, which still seems to be a distant goal), but also as a tool for music creators, allowing them to easily incorporate new styles and ideas into their work.

In computer vision, the most popular task in this direction is \emph{style transfer}, where the algorithm has two inputs: the `content' image to transform and a `style' image, bearing the style that we wish to impose on (or \emph{transfer} to) the content image.
On the other hand, work done on music so far has mostly focused on a different task, which we refer to as \emph{style translation}.
Contrary to style transfer, only the `content' input is given, and the goal is to render it in a target style which is known in advance and usually \emph{learned} from a large set of examples.
Note that although this second task is often also referred to as `style transfer' in the context of music and text generation, we claim that this conflicts with how the term is traditionally understood \cite{Efros2001ImageQF,Xie2007FeatureGT,Gatys2016ImageNetworks},
and that the term `translation' is more appropriate and in line with other prior work \cite{Isola2016Image-to-ImageNetworks,Zhu2017,Malik2017NeuralStyle,Mor2018}.

The focus of our work is on the latter task, and more specifically, on \emph{accompaniment style translation} for \emph{symbolic music}.
In particular, given a piece of music in a symbolic representation, our goal is to generate a new accompaniment for it in a different arrangement style while preserving the original harmonic structure. Even though our approach is generic, to narrow down our scope, we focus on generating bass and piano tracks.

A major difficulty of the music style translation task is that there are no publicly available `aligned' or `parallel' datasets (containing examples of the same music played in different styles). As a result, recent works closely related to ours
\cite{Brunner2018,Brunner2018a}
have adopted unsupervised learning frameworks~-- variational autoencoders (VAE) \cite{Kingma2014AutoEncodingVB} and CycleGANs \cite{Zhu2017}~--
and applied them to genre-labeled datasets.
However, these extensions to symbolic music have not yet permitted to obtain results as compelling as those on images \cite{Liu2017UnsupervisedNetworks,Zhu2017}, text \cite{Lample2017UnsupervisedOnly,Zhao2018AdversariallyRA}, and music audio \cite{Mor2018}.

In this study, we adopt a different strategy to overcome the lack of aligned data, which is to \emph{synthesize} it.
Synthetic training data has proven useful for music information retrieval tasks such as chord recognition \cite{Lee2008AcousticCT} and fundamental frequency estimation \cite{Mauch2014PYINAF,Salamon2017AnAF}, and is also popular for tasks like semantic segmentation 
in computer vision \cite{Ros2016TheSD,varol2017learning}.
In our case, synthetic data opens up the possibility for supervised learning techniques known from the machine translation field.
Moreover, it allows us to work with fine-grained style labels, as opposed to genre labels, which may be too vague or ambiguous
for such purposes.

Our main contributions are as follows:
\begin{compactitem}
    \item We propose a supervised, end-to-end neural model for symbolic music style translation, along with a training data generation scheme.
    \item Our model is able to translate into a large number of different styles by conditioning a single decoder on the target style. To our knowledge, this is the first time this technique has been applied to music translation with some success.
    \item To evaluate the performance of our model, we propose an objective metric of music style similarity.
    \item We show that an approach to music style translation based entirely on synthetic data is viable and generalizes well to more `natural' inputs, even in unrelated styles.
\end{compactitem}
We believe that our approach will foster new directions in this line of research; some of these will be briefly discussed in \hyperref[sec:conc]{the conclusion}. The source code of our system, built using TensorFlow, is available online.\footurl{https://git.io/musicstyle}

\section{Related work}
The work performed so far in the area of music style transformation is relatively small in volume but fairly diverse, since, as noted in \cite{Dai2018MusicPaper}, the transformations can work with different music representations as well as on different conceptual levels.

To our knowledge, the only work on music style transfer~-- in the original sense, as discussed in the \hyperref[sec:introduction]{introduction}~-- has been done on audio.
Some approaches \cite{Driedger2015LetIB,Tralie2018CoverSS} combine signal decomposition techniques with \emph{musaicing} \cite{Zils2001MUSICALM} (a form of concatenative synthesis). 
In \cite{Grinstein2018AudioST}, the authors attempt to transfer `sound textures' from a recording by means of techniques adapted from image style transfer, but without specific focus on the musical aspects.
In both cases, the transformation is largely limited to timbre.

The problem of unsupervised music audio translation is tackled in \cite{Mor2018}, where
the authors train a neural network to translate between a number of domains.
For symbolic music, style translation is studied in \cite{Brunner2018,Brunner2018a}, adapting unsupervised learning techniques from computer vision.
A different approach is proposed in \cite{Lu2018TransferringModels}, consisting in training a model on the target style only and then using pseudo-Gibbs sampling to transform a given piece of music. %

Finally, we should mention more `constrained' problems from the symbolic music domain which can also be framed as style translation tasks, e.g.\ (re\nobreakdash-)harmonization \cite{Pachet2014,Hadjeres2016StyleIA} and expressive performance generation \cite{Widmer2009YQXPC,Flossmann2011TowardAM,Malik2017NeuralStyle}.

\section{Synthetic data generation}
\label{sec:data}

\begin{figure*}
    \centering%
    \includegraphics[height=2.3cm,width=0.99\linewidth]{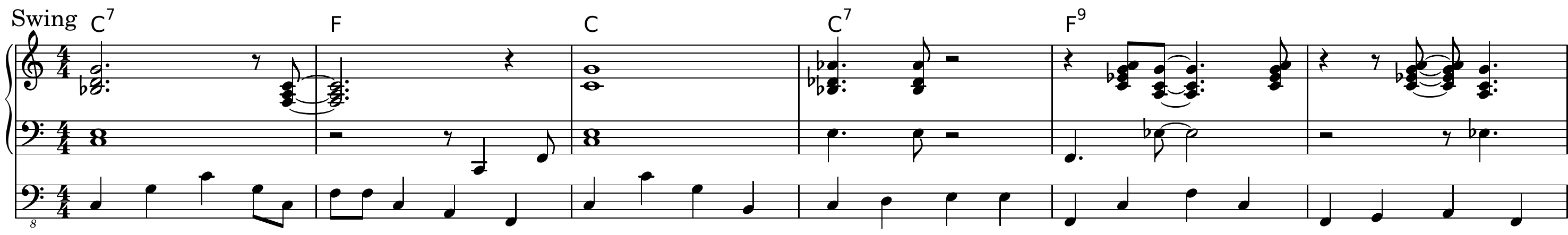}\\%
    \includegraphics[height=2.3cm,width=0.99\linewidth]{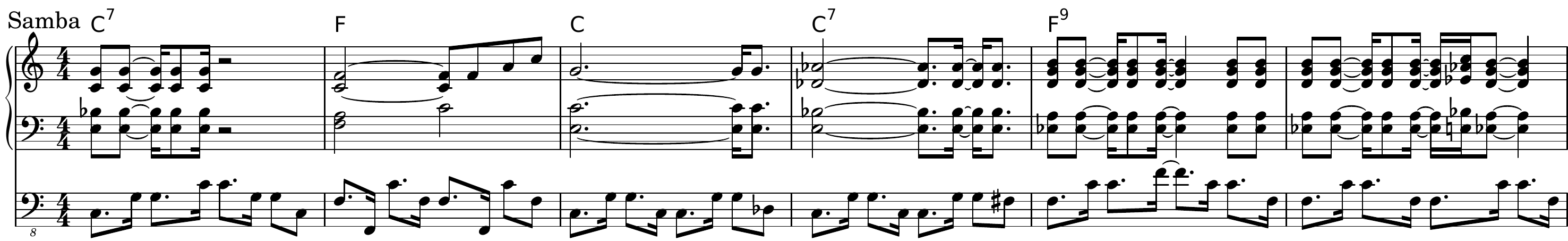}
    \caption{Six bars of an accompaniment (piano and bass) for a 12-bar blues,
    generated using BIAB in a `jazz swing' style (top) and a `samba' style (bottom).
    The timing is only approximate.
    The input chord sequence is displayed at the top.}
    \label{fig:biab-example}
\end{figure*}

Since we are in a supervised setting, our approach requires a large amount of \emph{paired} examples where each pair consists of one musical fragment arranged in two different styles.
Given that no such dataset is currently available, we created a synthetic one, generated using RealBand from the Band-in-a-Box (BIAB) software package \cite{band-in-a-box}.

First, we downloaded chord charts of around 3.5K songs in the BIAB format from a popular online archive \cite{biab-file-archive}.
We used BIAB to generate arrangements of these songs in different styles and filtered the resulting MIDI files to keep only those in \timesig{4}{4} or \timesig{12}{8} time.\footnote{The time signature depends on the style as well as on the song itself. A song originally in \timesig{4}{4} may have a \timesig{12}{8} arrangement and vice versa.}
We then chopped those files into segments of 8 bars, splitting notes that overlap segment boundaries.

We selected a total of 70 styles from the `0~MIDI' and `1~MIDI' style packs included in Band-in-a-Box 2018, representing a wide variety of popular music genres. Each style contains up to 5 accompaniment tracks (drums, bass, piano, guitar, strings).%
\footnote{These 5 labels are not always accurate; for example, some styles have two guitar tracks, one of which is labeled as piano.}
We generated each song in 3 randomly picked styles, providing $2\times\binom{3}{2}=6$ training pairs per segment, or around 658K training examples in total.
An example of a possible training pair is shown in \cref{fig:biab-example}.

In all experiments, we used 2,809 songs for training, 46 songs as a validation set and 46 songs for evaluation, each in 3 examples in different styles. The song names, along with the styles used for each song, are included in the supplementary material \cite{ismir2019-supplementary}.

\section{Proposed model}
We propose an architecture based on RNN encoder-decoder sequence-to-sequence models with attention \cite{Bahdanau2014NeuralTranslate}, commonly employed in machine translation and other areas of natural language processing.
This choice is motivated by the successes of RNNs on symbolic music generation \cite{Eck2002FindingTS,performance-rnn-2017,pmlr-v70-hadjeres17a,Sturm2016MusicTM} and by the ability of the attention mechanism to condition the generation on arbitrary input data without a prior alignment.

Our model is designed so that it is capable of translating music between a potentially large number of different styles.
This is achieved by conditioning the decoder on the target style.
An obvious advantage of this design is efficiency: to translate between $n$ styles, we only need to train a single model, compared to $n$ models (one for each target style; possibly with a shared encoder as in \cite{Mor2018}) or even $\Theta(n^2)$ models (one for each pair of styles, e.g.\ \cite{Brunner2018,Brunner2018a}).
Other implications of this choice are investigated in \cref{sec:one-to-one}.

On the other hand, to simplify the task and facilitate evaluation, we train a dedicated model for each target instrument track.
Our output representation and decoder architecture are chosen accordingly and would not necessarily be suitable for generating several independent tracks.
\vspace{10pt}
\noindent \textbf{Input and output representation. \hspace{3pt}}
A common choice of representation of symbolic non-monophonic music for neural processing is a piano roll. We use a binary-valued piano roll with 128 pitches and 4 columns per beat (quarter note) to encode our input.

\begin{figure}
    \centering
    \begin{minipage}{.85\linewidth}
    \begin{minipage}[c]{.55\linewidth}
    \includegraphics[width=\linewidth,trim={0 0 0 0.2cm},clip]{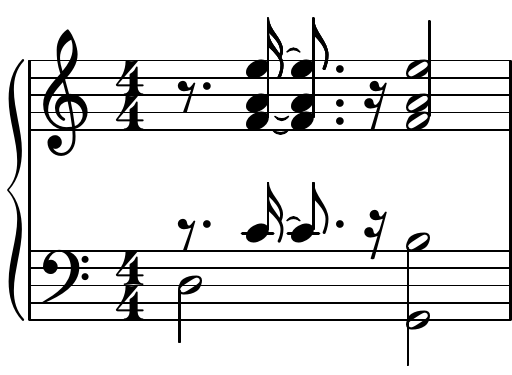}
    \end{minipage}%
    \hfill%
    \begin{minipage}[c]{.42\linewidth}
    \includegraphics[width=\linewidth]{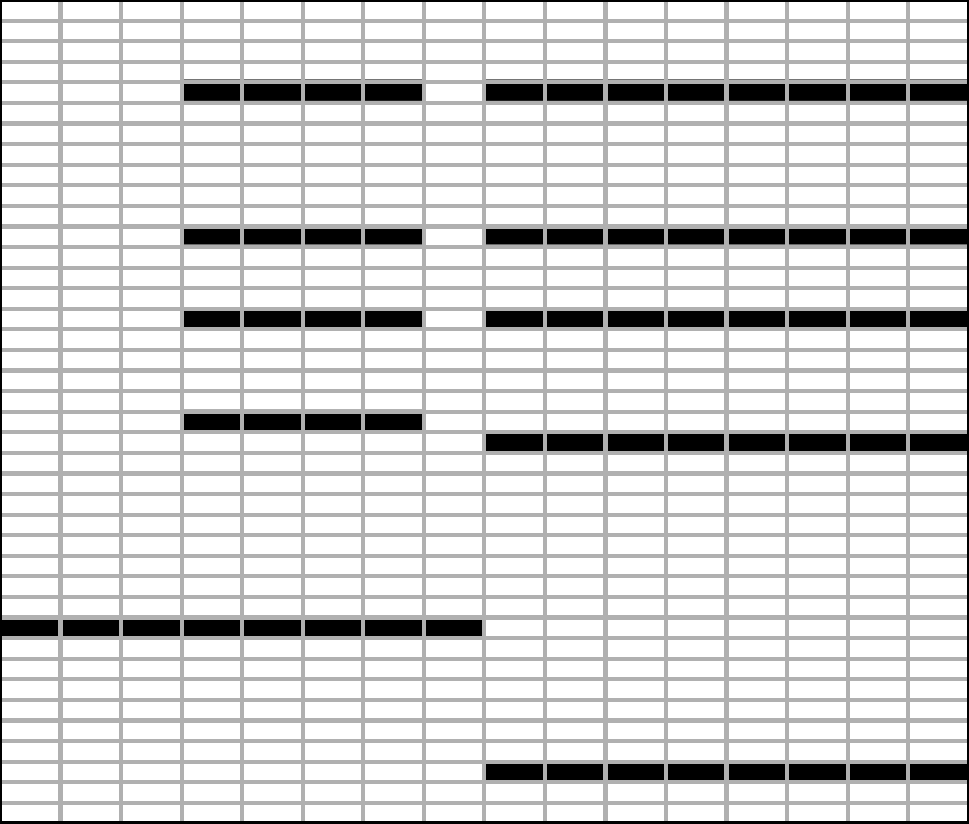}
    \end{minipage}\\%
    \begin{minipage}{\linewidth}
    \fontsize{6.7}{6}\selectfont%
    \texttt{%
    NoteOn(50)
    TimeShift(9)
    NoteOn(60)
    NoteOn(65)
    NoteOn(69)
    NoteOn(76)
    TimeShift(12)
    NoteOff(60)
    NoteOff(65)
    NoteOff(69)
    NoteOff(76)
    TimeShift(3)
    NoteOff(50)
    NoteOn(43)
    NoteOn(59)
    NoteOn(65)
    NoteOn(69)
    NoteOn(76)
    TimeShift(24)
    NoteOff(All)
    }
    \end{minipage}
    \end{minipage}
    \caption{A bar of music, represented as a piano roll (top right) and as a sequence of 20 event tokens (bottom).}
    \label{fig:encoding}
\end{figure}

For representing the output (and also as an alternative input representation), we opted for a MIDI-like encoding, which~-- unlike a piano roll~-- is straightforward to model using an RNN decoder.
Specifically, following \cite{performance-rnn-2017}, we encode the music as a sequence of 3 types of events, each with one integer argument:
\begin{itemize}[nosep,leftmargin=1.5em,labelwidth=0.4em,align=left]
    \item \texttt{NoteOn(\textit{pitch})}: start a new note at the given pitch;
    \item \texttt{NoteOff(\textit{pitch})}: end the note at the given pitch;
    \item \texttt{TimeShift(\textit{delta})}: move forward in time by the specified amount, measured in 12ths of a beat.
\end{itemize}
\texttt{NoteOn} and \texttt{NoteOff} take values in the range 0--127, whereas \texttt{TimeShift} is within 1--24.\footnote{When encoding the piano track, we compress the sequences by also including a \texttt{NoteOff(All)} event which ends all currently active notes.}
In contrast to \cite{performance-rnn-2017}, our representation is tempo-invariant and we do not model dynamics.
\cref{fig:encoding} illustrates both representations.

\vspace{10pt}
\noindent \textbf{Model architecture and training. \hspace{3pt}}
\label{sec:model_arch}
The proposed model consists of an encoder and a decoder; the former serves to compute a dense representation of the input, while the latter generates the output event sequence, conditioned on the encoded input and the target style.

The architecture of the encoder depends on the type of input representation:
\begin{compactitem}
\item If the input is a piano roll, we use a two-layer convolutional network (CNN),
followed by a bidirectional RNN with a gated recurrent unit (GRU) \cite{Cho2014LearningPR}.
The CNN serves to compress the input, resulting in a sequence of 1280-dimensional vectors with 2 vectors per bar.
The bidirectional GRU then adds the ability to incorporate information from a wider context.
\item If the input is a sequence of tokens, we use an embedding layer, also followed by a bidirectional GRU.\vspace{3pt}
\end{compactitem}
We refer to the two variants of the model as `roll2seq' and `seq2seq', respectively.

\begin{figure}
    \centering
    \includegraphics[scale=0.85]{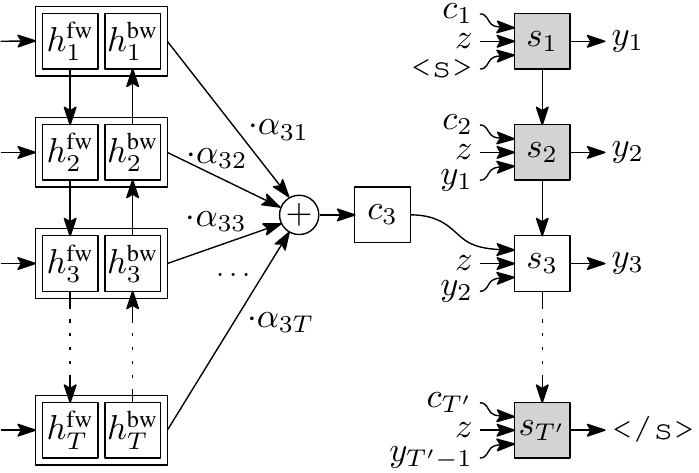}
    \caption{The attention-based decoder. 
    During the $i$-th decoding step (here $i=3$), a set of coefficients $\alpha_{ij}$ is computed and used to weight the encoder states $h_j=[h^{\mbox{\scriptsize fw}}_j,h^{\mbox{\scriptsize bw}}_j]$ to obtain the context vector $c_i$, which in turn is used as input for the decoder cell to compute the next state, $s_i$.}
    \label{fig:attention}
\end{figure}

The decoder is also implemented using a GRU, conditioned on the target style and equipped with a feed-forward attention mechanism \cite{Bahdanau2014NeuralTranslate} acting on the encoder outputs.
More precisely, as illustrated in \cref{fig:attention}, the $i$-th decoder state $s_i$ is computed as
\begin{equation*}
s_i = \text{GRU}([c_i,W^\mathrm{s} z,W^\mathrm{e} y_{i-1}], s_{i-1}),
\end{equation*}
where $[\cdot]$ denotes concatenation, $z$ and $y_{i-1}$ respectively denote the one-hot encoded representations of the target style and the previous output event,
$W^\mathrm{s},W^\mathrm{e}$ are the corresponding embedding matrices,
and $c_i$ is the context vector.
The latter %
is a weighted average of the encoder outputs, computed by the attention mechanism.
The purpose of attention is to provide an \emph{alignment} between the encoder and decoder states.
The need for this alignment arises from the fact that the positions in the output sequence are not linear in time (due to the chosen encoding), and the decoder therefore needs to be able to move its focus flexibly over the input.
For a complete description of attention, see \cite{Bahdanau2014NeuralTranslate}.

\begin{figure}
    \centering
    % \documentclass{minimal}
% \usepackage{tikz}
% \usetikzlibrary{calc,positioning,arrows,decorations.text} 
% \begin{document}
{
\small
\begin{tikzpicture}[x=0.8cm,y=0.7cm,auto]
    \tikzset{
        %Define standard arrow tip
        >=stealth',
        %Define style for boxes
        module/.style={
           rectangle,
           rounded corners,
           draw=black, thick,
           text width=6.5em,
           minimum height=2em,
           text centered},
        data/.style={
           rectangle,
           draw=black,
           minimum height=1.5em,
           inner xsep=0.5em,
           inner ysep=0em,
           text centered},
    }
    \pgfdeclarelayer{back}
    \pgfsetlayers{back,main}

    \node[module] (biab) {BIAB};
    \node[data,above=0.9 of biab] (chart) {chord chart}
        edge[->] (biab.north);
    \node[right=3.2 of biab] (dummy) {};
    \node[module,above=0.4 of dummy] (enc) {encoder};
    \node[module,below=0.4 of dummy] (dec) {decoder}
        edge[<-] (enc.south);
    \node[data,above=0.75 of enc] (src) {source $x$}
        edge[->] (enc.north);
    \node[data,below=0.75 of dec] (tgt) {target $y$}
        edge[<-,densely dashed] %node[right] {$\mathcal{L}$}
        (dec.south);
    \draw[->]
        ([yshift=0.1cm]biab.east) to %[out=20,in=-150]
        node [midway,above=0,sloped] {source style}
        (src.west);
    \draw[->] 
        ([yshift=-0.1cm]biab.east) to %[out=-20,in=150]
        node [midway,below=0,sloped] {target style $z$}
        (tgt.west);
    \node[data,right=0.5 of dummy] (z) {$z$};
    \draw[->] (z.west) -| ([xshift=0.2cm]dec.north);
    
    \begin{pgfonlayer}{back}
    \draw[dotted,darkgray] ($(biab)!0.5!(dummy) + (0, -2.9)$) -- ($(biab)!0.5!(dummy) + (0, 3.7)$);
    \node at ($(biab) + (0, 3.5)$) {(a)\enskip\textbf{Data generation}};
    \node at ($(dummy) + (0, 3.5)$) {(b)\enskip\textbf{Training}};
    \end{pgfonlayer}
\end{tikzpicture}
}
% \end{document}
    {\phantomsubcaption\label{fig:pipeline-data}}%
    {\phantomsubcaption\label{fig:pipeline-training}}%
    \caption{A scheme of the training pipeline.
        \protect\begin{inparaenum}[(a)]
            \protect\item We use BIAB to generate each song in different arrangement styles (see \cref{sec:data}). 
            \protect\item The model is trained to predict the target-style segment $y$ given a source segment $x$ and the target style $z$ (see \cref{sec:model_arch}).
        \protect\end{inparaenum}%
    }
    \label{fig:pipeline}
\end{figure}
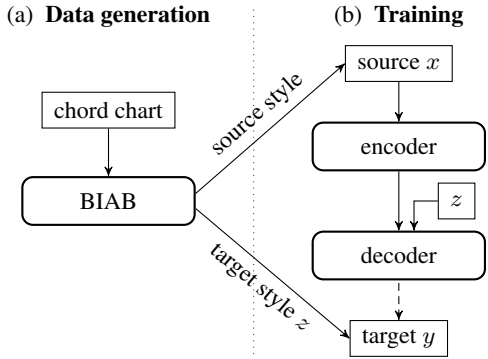

The training pipeline is portrayed in \cref{fig:pipeline-training}. Each training example consists of a song segment $x$ in one style (the source style) along with the corresponding segment $y$ in a different style ($z$, the target style). We train the model by minimizing the loss on $y$ while passing $x$ to the encoder and conditioning the decoder on $z$.

The models are trained using Adam \cite{Kingma2015AdamAM} with learning rate decay and with early stopping on the development set.
Our configuration files with complete hyperparameter settings are included with the source code.

Once the model is trained, we perform style translation using greedy decoding, i.e.\ by taking the most likely output token at every step (and using that as input in the next step). We also explored random sampling with different softmax temperatures, but found that this leads to a higher number of errors (i.e.\ invalid sequences or incorrect timing) and does not significantly improve the quality of the outputs.

\section{Evaluation metrics}
\label{sec:metrics}
When evaluating a style transformation, we need to consider two complementary criteria: how well the transformed music fits the desired style (\emph{style fit}) and how much content it retains from the original (\emph{content preservation}).
Note that it is trivial (but useless) to achieve perfect results on either of these two criteria \emph{alone}, so it is essential to evaluate both of them.

In this section, we describe `objective', automatically computed metrics for both criteria.
Even though we believe these metrics are sound and well-motivated, we acknowledge the limitations of automatic metrics in general and encourage the reader to listen to the provided example outputs \cite{ismir2019-supplementary} to get a real sense of their quality.

\vspace{10pt}
\noindent \textbf{Content preservation. \hspace{3pt}}
We use a content preservation metric similar to the one proposed by \cite{Lu2018TransferringModels}, computed by correlating the chroma representation of the generated segment with that of the corresponding segment in the source style.
This is motivated by the fact that we expect the output to follow the same sequence of chords as the input.
More precisely, we compute chroma features for each segment at a rate of 12 frames per beat and smooth each of them using an averaging filter with a window size of 2 beats (24 frames) and a stride of 1 beat (12 frames).
Finally, we calculate the average frame-wise cosine similarity between the two sets of chroma features.

\vspace{10pt}
\noindent \textbf{Style fit. \hspace{3pt}}
In some of the recent music style transformation works \cite{Brunner2018,Brunner2018a}, the quality of a transformation is measured by means of a binary style classifier trained on a pair of styles.
However, the merit of such evaluation is limited, since a high classifier score merely demonstrates that the output has some of the distinguishing features of the target style, and not necessarily that it actually fits the style.
For this reason, we aim for a more interpretable metric of style fit.

As observed by \cite{McKay2004AutomaticGC,Hadjeres2016StyleIA,Sakellariou2017MaximumEM}, musical style is well captured in pairwise statistics between neighboring events. Drawing inspiration from the features proposed in \cite{McKay2004AutomaticGC}, we devise 
a key- and time-invariant style representation which we call the \emph{style profile}.

To compute the style profile, we consider all pairs of note onsets less than 4 beats apart and at most 20 semitones apart, and record the time difference and interval for each pair. In other words, we define the following multiset of ordered pairs:
\begin{equation*}
\begin{aligned}
    \mathcal{S} = \{ (t_b-t_a, \> p_b-p_a) \,|\, a,b\in \text{notes},\ a\neq b, \\
    0 \leq t_b - t_a < 4, \> \abs{p_b-p_a} \leq 20\},
\end{aligned}
\end{equation*}
where $t_x$ is the onset time of the note $x$ (measured in fractional beats) and $p_x$ is its MIDI note number.
We then obtain the style profile as a normalized 2D histogram of $\mathcal{S}$ with 6 bins per beat and one bin per semitone, and flatten it to get a 984-dimensional vector.

Finally, to quantify the style fit of a particular set of outputs, we compute their style profile and measure its cosine similarity to a reference profile.
Note that an 8-bar segment may not be sufficient to obtain a reliable style profile; instead, we always aggregate the statistics over a number of segments.
In particular, we put forward two variants of the style fit metric, obtained as follows:
\begin{compactenum}[(a)]
\item Compute a style profile aggregated over all outputs of a model in a given target style and measure its cosine similarity to the reference.
\item Compute a style profile for each translated song separately and measure its cosine similarity to the reference.
We report the mean and standard deviation over all songs.
\end{compactenum}
We refer to (a) and (b) as `macro-style' and `song-style', respectively.
In both cases, the reference style profile is extracted from the training set, separately for each track.

\begin{figure}
    \centering
    \includegraphics[width=\linewidth]{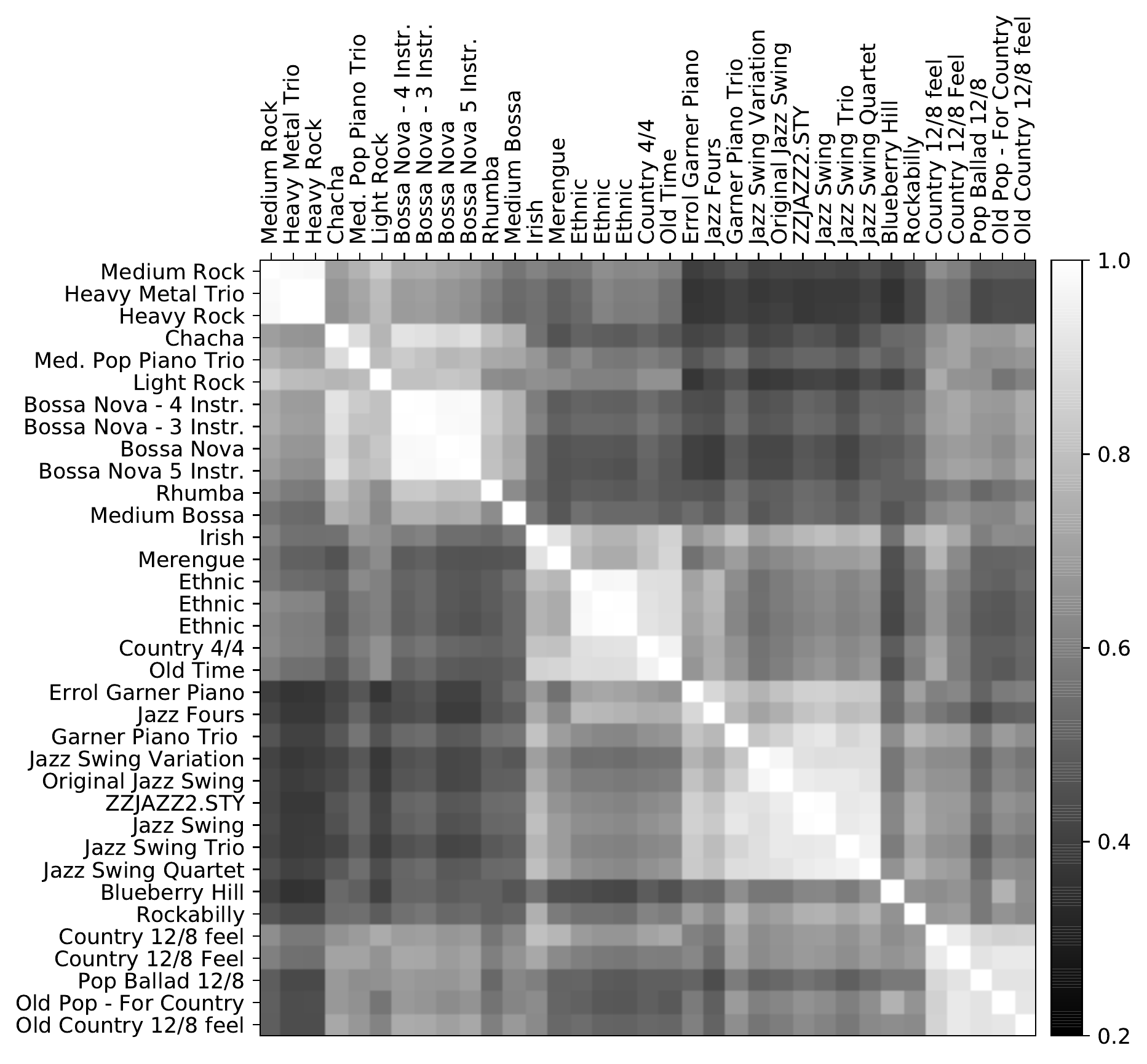}
    \caption{Pairwise cosine similarities of selected style profiles computed on training bass tracks. The styles are ordered based on a hierarchical clustering of the profiles.}
    \label{fig:profile_similarities}
\end{figure}

While we do not claim that this metric is able to distinguish between broad style categories (such as genres), yet it can definitely capture the differences and similarities between specific `grooves', which makes it well-suited for our purpose.
This is illustrated in \cref{fig:profile_similarities}, showing the pairwise similarities between the profiles of the bass tracks of different BIAB styles, with clearly visible clusters of jazz, rock or country styles. %

\begin{figure*}
    \centering
    \begin{subfigure}[b]{0.495\linewidth}
    \includegraphics[width=\linewidth]{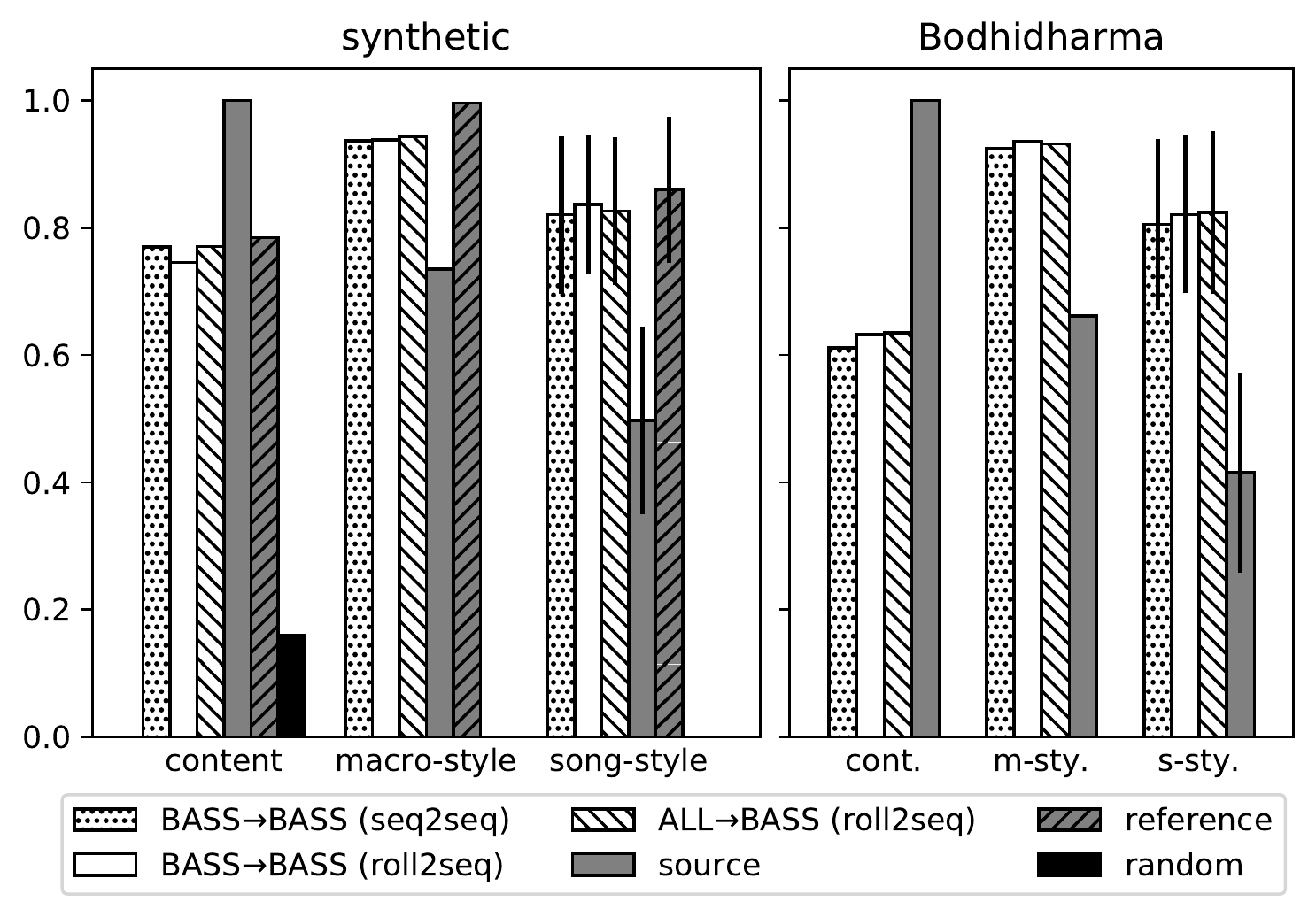}
    \caption{\track{bass}}
    \label{fig:bass_results}
    \end{subfigure}
    \hfill
    \begin{subfigure}[b]{0.495\linewidth}
    \includegraphics[width=\linewidth]{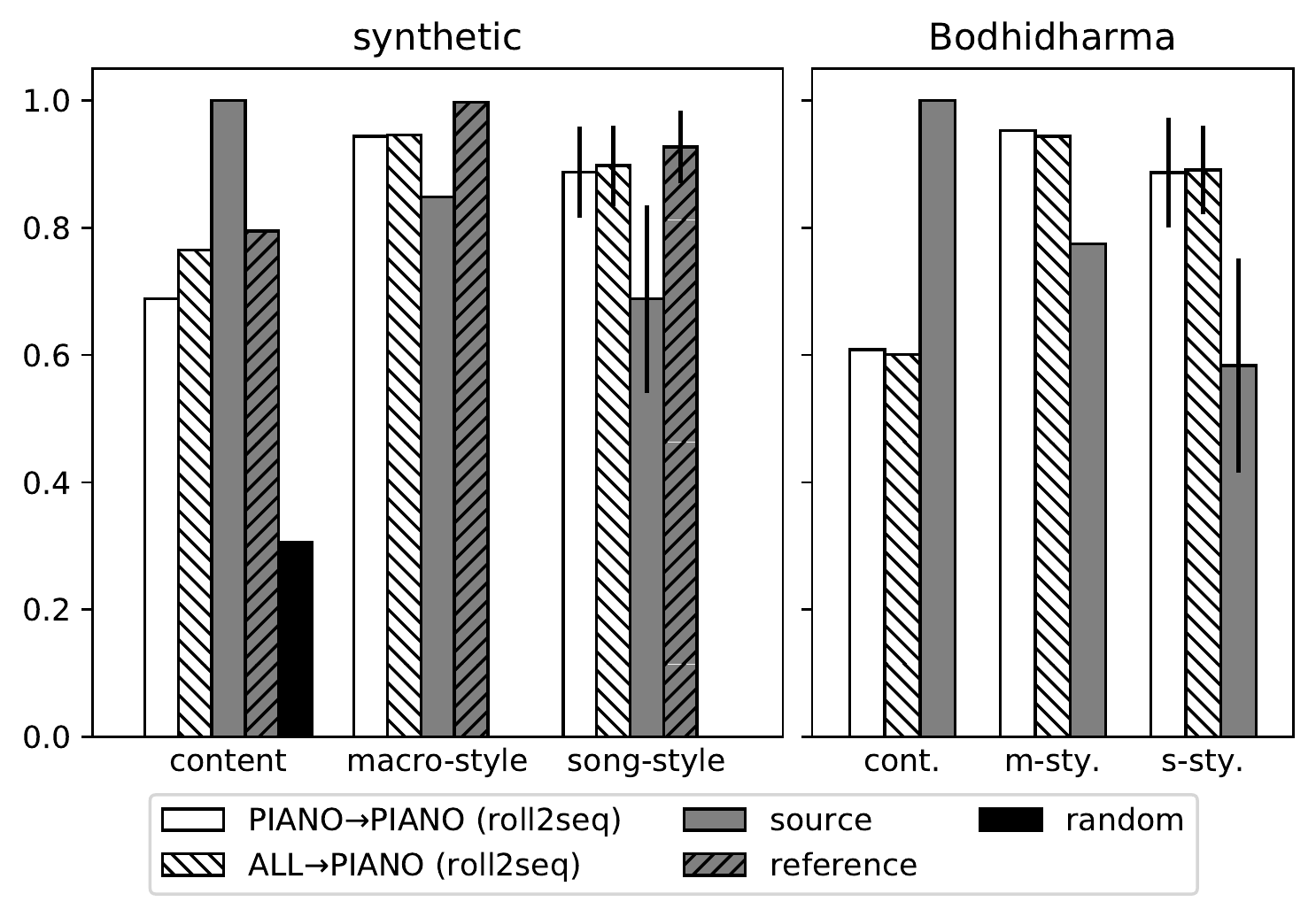}
    \caption{\track{piano}}
    \label{fig:piano_results}
    \end{subfigure}
    \caption{Evaluation results on content preservation and style fit. `Source' is the original track (bass or piano), `reference' is a track generated by BIAB in the target style and `random' is a random permutation of the references. For `song-style', we plot the mean and the standard deviation over all songs and target styles.}
    \label{fig:results}
\end{figure*}

\section{Experimental results}
In our experiments, we focus on generating the bass and piano tracks, and we train a dedicated model for each of them.
For each track, we consider two scenarios: generating the track given only the corresponding source track (\trackpair{bass}{bass}, \trackpair{piano}{piano}), and using all non-drum accompaniment tracks from the input (\trackpair{all}{bass}, \trackpair{all}{piano}).

For \trackpair{bass}{bass}, we compare the seq2seq and roll2seq architectures defined in \cref{sec:model_arch}.
For all other pairs, where the input is non-monophonic, we only employ roll2seq, since the sequential representation grows disproportionately in length in these cases and the computational cost of the attention mechanism becomes too heavy.

We evaluate our models on our synthetic test set generated by BIAB and on the
Bodhidharma MIDI dataset \cite{McKay2005TheBS}.
The latter is a diverse collection of 950 MIDI recordings annotated with genre labels. %
We filtered and pre-processed the dataset in the same way as the synthetic test set and we extracted the bass and piano tracks.\footnote{To form the bass track, we retrieve all notes assigned to any Bass instrument. For the piano track, we use the Piano and Organ classes.}

We also made extensive attempts to train the recent models of \cite{Brunner2018,Brunner2018a,Lu2018TransferringModels} on our data using the source code published by the authors, but unfortunately without success. This has prevented us from comparing these models with our proposal. Nonetheless, the provided outputs \cite{ismir2019-supplementary} can serve as a basis for perceptual comparison.

\subsection{Evaluation}
For a comprehensive evaluation of each model, we translated all inputs to all 70 styles and calculated the content preservation and style fit metrics. The results (averaged) are presented in \cref{fig:results}.

We provide two baselines for each track (bass and piano): `source', which is simply the same track before the translation, and `reference', which is a track generated by BIAB based on the chord chart (only available for the synthetic test set).
As expected, the style fit is low for the source track (measured with respect to the target style) and close to 1 for the reference track.
Our models' outputs generally do not fit the target style as perfectly as the reference does, but still score high compared to the source.

As for content preservation, we can notice that the reference value is quite low (0.78 for \track{bass} and 0.79 for \track{piano}).
This should not be too surprising, since we are comparing accompaniments in two different styles, which might have different pitch-class distributions; moreover, there is some random harmonic variation within each style (see e.g.\ bars 5--6 in \cref{fig:biab-example}).
The results achieved by our models on the synthetic test set are very close to the reference.
To illustrate the value range of the metric, we provide the results obtained by a `randomized' baseline (shown as `random' in \cref{fig:results}), where we randomly permuted the reference segments for each style (obtaining a reference with the correct style, but the wrong content).
The resulting value is very low (0.16 for \track{bass} and 0.31 for \track{piano}) compared both to the true reference and to our models, indicating that the metric is useful and the models are performing well.

On Bodhidharma, content preservation is generally weaker than on the synthetic test set.
One interpretation can be that the encoder simply fails to extract the content information accurately, since it was trained on a different domain.
However, we also find that the models often make timing errors on Bodhidharma inputs, leading to misalignment between the input and the output, which may also cause the content preservation metric to drop.

On the other hand, the style fit on Bodhidharma is close to the results on the synthetic test set (and not consistently lower or higher), and the difference to `source' (i.e.\ the corresponding input track) is more marked, perhaps reflecting a higher style variability in the Bodhidharma data.

Upon listening, we clearly observe that the outputs are musical and seem to both fit the target style and follow the harmonic structure of the inputs. 
Besides, even though the piano and the bass tracks are generated independently, they sound surprisingly coherent. However, as mentioned above, we also observe occasional timing errors (especially in heavily syncopated grooves), which become more prominent when the bass and piano tracks are combined. A potential remedy for this issue would be to modify the encoding to make it more robust, e.g.\ by representing the timing in a beat-aware manner.
We also note that the single-track models output harmonically incorrect notes more often than the \track{all} models; this is expected, since their \emph{input} is less harmonically rich.
This effect is clearly audible (especially in \track{bass}, where important scale degrees are often missing in the input), but cannot be captured by the content preservation metric, which is computed against the same input.

\begin{figure}
    \centering
    \includegraphics[width=\linewidth,]{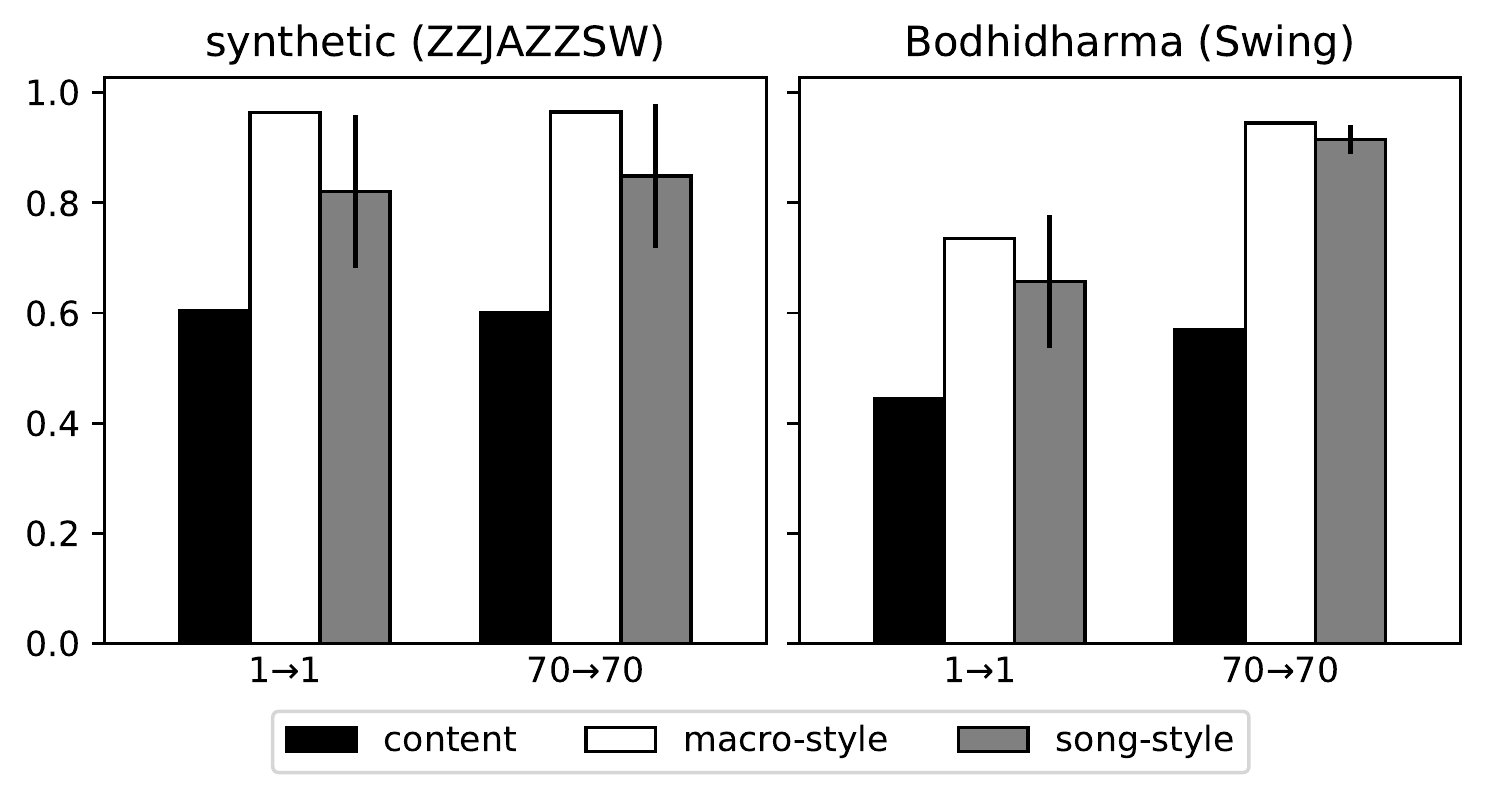}
    \caption{Comparison of a single-style-pair model (1$\to$1) and a full model (70$\to$70) on the ZZJAZZSW$\to$TWIST style pair.}
    \label{fig:bass_pair_eval}
\end{figure}

\subsection{Comparison with a single-pair model}
\label{sec:one-to-one}
All models presented so far were trained on music in 70 different styles, as opposed to a single style pair.
To investigate the effect of this choice, we picked a pair of fairly dissimilar styles~-- ZZJAZZSW (`Jazz Swing Variation') and TWIST (`Twist Style', categorized as `Lite Pop')~-- and generated a new training, validation and test set with each song rendered in these two styles only.
To increase the amount of data, we performed this twice for each song (with different results), obtaining $2\times2=4$ training pairs per segment.

We used this new dataset to train single-style-pair versions of all models (in the ZZJAZZSW$\to$TWIST direction only), preserving the original architectures except for the conditioning on the target style.
We compare these `1$\to$1' models to the full versions (70$\to$70) on two sets of inputs:
\begin{compactitem}
\item the synthetic test set in the ZZJAZZSW style;
\item the `Swing' section of Bodhidharma (23 songs).
\end{compactitem}

In \cref{fig:bass_pair_eval}, we show the results for the two variants of the \trackpair{all}{bass} model.
While the performance on the synthetic data seems to be the same, the scores of the 1$\to$1 model drop considerably on the Bodhidharma data, suggesting that the model is overfitted to the `synthetic' swing style.
On the other hand, the performance of the 70$\to$70 model stays high, showing that training on many different styles helped the model generalize to real swing.

\subsection{Style embedding analysis}

\begin{figure}
    \centering
    \includegraphics[width=0.8\linewidth]{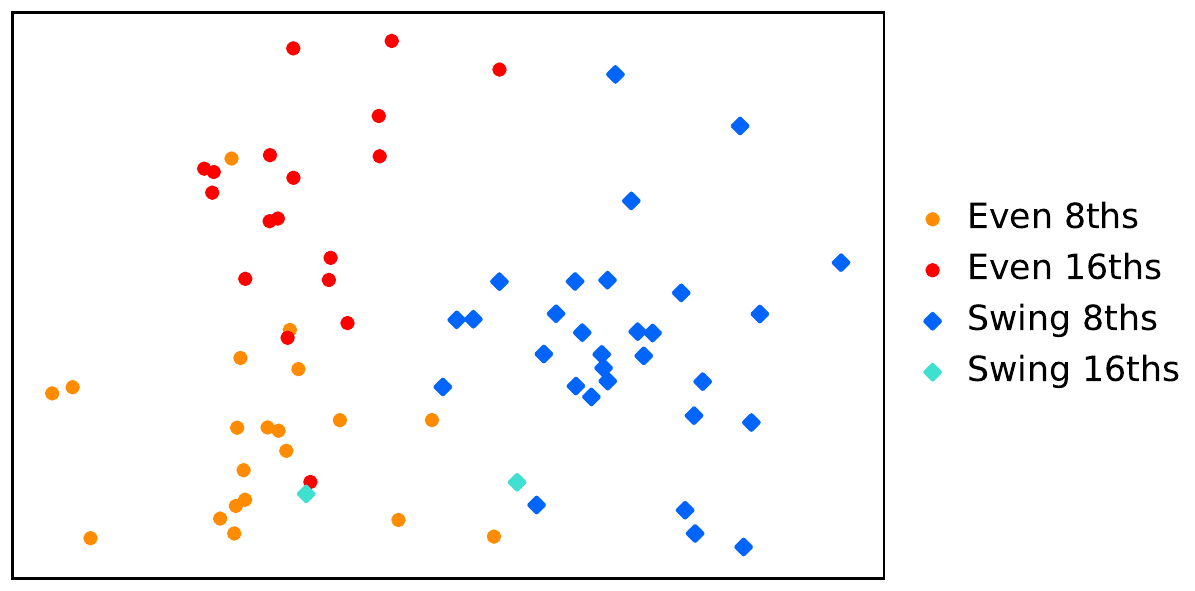}
    \caption{Style embeddings learned by the \trackpair{all}{piano} model, labeled with `feel' annotations provided by BIAB. Dimensionality reduction was performed using linear discriminant analysis (LDA) with the feel labels as targets.}%
    \label{fig:embedding_projection}
\end{figure}

Neural representation spaces are often found to exhibit a meaningful geometry, and our learned style embedding space is no exception.
As an example,  \cref{fig:embedding_projection} shows a projection of the embeddings labeled by the `feel' of each style, with `even' and `swing' feel styles being clearly separated.
We include more plots in the supplementary material and also make available an interactive visualization.%
\footurl{https://bit.ly/2G5Jgnq}

\section{Conclusion}
\label{sec:conc}
In this study, we focused on symbolic music accompaniment style translation.
As opposed to the current methods, which are inherently restricted to be unsupervised due to the lack of aligned datasets, we developed the first fully supervised algorithm for this task, leveraging the power of synthetic training data.
Our  experiments  show  that  our  models are  capable of producing musically meaningful accompaniments even for real MIDI recordings.

We believe that these results point to interesting research directions.
First, synthetic data seem to be an excellent resource for music style translation, and could be used as a starting point even for unsupervised learning, allowing to validate a given approach before moving on to more challenging, unaligned datasets.
Second, our supervised approach could be used to address more general music transformation tasks, and we are already working on an extension in this direction.

\section{Acknowledgement}
This research is supported by the European Union’s Horizon 2020 research and innovation programme under the Marie Skłodowska-Curie grant agreement No.\ 765068 (MIP-Frontiers) and by the French National Research Agency (ANR) as a part of the FBIMATRIX (ANR-16-CE23-0014) project.

\bibliography{ondrej_phd}

\end{document}